\documentclass[twocolumn]{aastex701}
\usepackage{xcolor}
\usepackage{subcaption}

\usepackage{amsmath}
\usepackage{tablefootnote}
\usepackage[section]{placeins}
\usepackage{comment}
\usepackage{hyperref}
\usepackage{makecell}
\usepackage{soul}

\begin{document}

\shorttitle{Effects of the Assumed Background on NICER Radius Estimates}
\shortauthors{Isiah M. Holt et al.}

\graphicspath{{./}{figures/}}

\title{An Investigation of Systematic Effects from Background Priors on PSR J0740$+$6620 Radius Estimates Using Synthetic NICER and XMM-Newton Data}
\author[0000-0002-3097-942X]{Isiah M. Holt}
\affiliation{Department of Astronomy, University of Maryland, College Park, MD 20742-2421, USA}
\affiliation{NASA Goddard Space Flight Center, Greenbelt, MD, USA}
\email[show]{imholt@umd.edu}

\author[0000-0002-2666-728X]{M. Coleman Miller}
\affiliation{Department of Astronomy, University of Maryland, College Park, MD 20742-2421, USA}
\affiliation{Joint Space-Science Institute, University of Maryland, College Park, MD 20742-2421, USA}
\email{mcmiller@umd.edu}

\author[0000-0001-6157-6722]{Alexander J. Dittmann}
\affiliation{Institute for Advanced Study, 1 Einstein Drive, Princeton, NJ 08540, USA}
\altaffiliation{NASA Einstein Fellow}
\email{dittmann@ias.edu}

\author[0000-0002-3862-7402]{Frederick K. Lamb}
\affiliation{Illinois Center for Advanced Studies of the Universe and Department of Physics, University of Illinois at Urbana-Champaign, 1110 West Green Street, Urbana, IL 61801-3080, USA}
\affiliation{Department of Astronomy, University of Illinois at Urbana-Champaign, 1002 West Green Street, Urbana, IL 61801-3074, USA}
\email{imholt@umd.edu}

\begin{abstract}

Accurate and precise measurements of neutron star radii provide invaluable information about the cold, dense matter in neutron star cores. Analyses of synthetic X-ray pulse waveform data for nonaccreting neutron stars of the type collected by the Neutron Star Interior Composition Explorer (NICER) have indicated that the mass and radius estimates extracted therefrom are robust against potential sources of systematic error, such as errors in the assumed pattern of the thermal X-ray emission from the surface of these stars. A potentially important but unexplored source of systematic error is misparameterization of unmodulated background components, which might bias the inferred radius, particularly when data from different telescopes are analyzed. In this study, we investigate the effects of the background model on radius estimates by jointly analyzing synthetic NICER and XMM-Newton data, using the $\sim 2.1~M_\odot$ pulsar PSR~J0740$+$6620 as a prototypical example. Our analysis shows that even if the background assumed in the model underestimates the actual background by a factor of more than 5, the resulting shift of the radius posterior from the true value of the radius corresponds to only $\sim1\sigma$. In all the cases we examined, the Bayesian evidence for the correct background model is greater than for the incorrect background model. These results add to the evidence that analyses of NICER-like data provide accurate measurements of radii when the statistical sampling is thorough and the model fits the data well.

\end{abstract}


\section{Introduction} \label{sec:intro}

Neutron star cores have a combination of high density, low temperature, and large neutron-proton asymmetry that cannot be reproduced in laboratories, and their properties cannot currently be predicted precisely using quantum chromodynamics (due to the fermion sign problem; see, e.g., \citealt{2005PhRvL..94q0201T,2010IJMPA..25...53H}). Consequently, observations of neutron stars provide uniquely valuable constraints on the properties of cold dense matter at densities up to a few times nuclear saturation density ($n_s\approx 0.16$~fm$^{-3}$), the highest density that occurs in nuclei when there is no confining pressure. Astrophysical observations of neutron stars are therefore crucial for understanding the nature of matter at supranuclear densities.

Most of the macroscopic properties of a neutron star, such as its maximum mass or its radius or tidal deformability as a function of mass, depend only on the equation of state (EOS) of the matter in its interior, i.e., the pressure as a function of the density, temperature, and other properties. In all but the youngest neutron stars, the thermal energies of the particles in this matter are much less than their Fermi energies. As a result, when this matter is in microscopic equilibrium, which is likely in neutron star cores, the pressure depends only on the energy density $\epsilon$, that is, $ P=P(\epsilon)$, where $P$ is the pressure.

Over the last $\sim15$ yr, numerous astronomical observations have helped constrain the EOS of cold, high-density matter. These include measurements of high masses for a few neutron stars such as PSR~J1614$-$2230 ($M = 1.97 \pm 0.04~M_\odot$, \citealt{2010Natur.467.1081D}), PSR~J0348$+$0432 ($M = 1.806 \pm.037~M_\odot$, \citealt{2024arXiv241202850S}), PSR~J0740$+$6620 ($M = 2.08 \pm 0.07~M_\odot$, \citealt{2021ApJ...915L..12F}), and PSR~J0952$-$0607 ($M = 2.35 \pm 0.17~M_\odot$, \citealt{2022ApJ...934L..17R}); observational constraints on the tidal deformability of neutron stars inferred from observations of the gravitational wave event GW170817 (see \citealt{2016PhRvL.116f1102A} and \citealt{2018PhRvL.121i1102D}); and estimates of the radii of several neutron stars using various approaches to analyzing thermal X-ray emission from their surfaces.

Early attempts to measure the radii of neutron stars in quiescent low-mass X-ray binary systems (\citealt{2011ApJ...732...88G,2013ApJ...772....7G,2014ApJ...784..123L,2016ApJ...831..184B,2018MNRAS.479.3634M,2021A&A...650A.139K}) and/or thermonuclear X-ray burst sources (see e.g., \citealt{1979ApJ...234..609V,2006astro.ph..1076M,2010ApJ...719.1807G,2010ApJ...722...33S,2013ApJ...776...19L,2016ApJ...820...28O}) used integrated flux and spectral observations.   However, radius estimates made using these methods are subject to significant systematic errors: though the models used in these analyses fit the data well (with some exceptions; see \citealt{2010ApJ...722...33S,2014MNRAS.445.4218K}), making different assumptions about the composition of a neutron star's atmosphere can change its inferred radius by $\sim 50$\%, far more than the apparent uncertainty of the measurement (\citealt{2013arXiv1312.0029M}). The usual assumption that the neutron star surface radiates uniformly can also introduce large systematic errors. By relaxing this assumption, \citet{2017A&A...608A..31N} obtained radii $\sim 10{\rm \%}-20\%$ larger than the radii they inferred using a model that assumed the surface temperature is uniform.

Earlier theoretical work had established the framework for modeling pulse profiles from neutron stars with hot polar caps, including the effects of rapid rotation, Doppler boosting, light travel time delays, and frame dragging \citep{2000ApJ...531..447B}. \citet{2001ApJ...546.1098W} provided the first survey of how waveform observables depend on the parameters now central to modern pulse profile modeling, including stellar compactness, spot size and location, and observer inclination, among others.

Another early approach attempted to model the X-ray pulsations of accretion-powered X-ray pulsars and then use the effects of the relativistic Doppler shift and aberration and gravitational lensing on the pulse profile in these models to infer the mass and radius of the star (see, e.g., \citealt{2003MNRAS.343.1301P}). This approach turned out to be unsatisfactory, because there are so many potential sources of significant systematic errors (see, e.g., \citealt{2008ApJ...675.1468H,2016EPJA...52...63M}). As a result, it was not possible to fully investigate the possible systematic errors to determine their sizes and nature of their effects. Consequently, the mass and radius estimates made using this method may have been susceptible to significant systematic errors. A detailed discussion of the history of using thermal hot spot modeling to infer neutron star masses and radii, including the approach that was ultimately realized using NICER, is given in Section 5 of \citet{2016EPJA...52...63M}.

These difficulties motivated searches for alternative methods to determine the masses and radii of neutron stars, which led to the proposal to determine the masses and radii of certain rotating neutron stars by accurately measuring the periodic pulse waveforms of the thermal X-ray emission from their surfaces and then fitting theoretical models to them (\citealt{2013ApJ...776...19L,2015ApJ...808...31M,2016EPJA...52...63M}). Analyses of synthetic pulse waveform data demonstrated that this method could be used to estimate the masses and radii of these stars with interesting precisions, and furthermore, that the resulting mass and radius estimates are not significantly biased by certain types of systematic errors, such as using incorrect shapes and temperature distributions for the X-ray emitting regions on the stellar surface, provided that the fit to the pulse waveform data is statistically acceptable (again see \citealt{2013ApJ...776...19L,2015ApJ...808...31M,2016EPJA...52...63M}). 

The Neutron Star Interior Composition Explorer (NICER; \citealt{2012SPIE.8443E..13G}), with its ability to measure the X-ray pulse waveforms of some nonaccreting neutron stars with high temporal and spectral resolution, made it possible to use the pulse waveform fitting method to determine the masses and radii of these neutron stars with unprecedented precision. The rms uncertainties in the photon arrival times measured using NICER are less than 100 ns, much less than the minimum observed rotational periods of neutron stars. NICER made it possible to accurately measure the X-ray flux and spectrum of these rotating neutron stars as a function of the star's rotational phase rather than merely their time-averaged values.

As noted above, the first studies of the measurements of neutron star masses and radii that could be achieved using data of this quality (\citealt{2013ApJ...776...19L,2015ApJ...808...31M,2016EPJA...52...63M}) demonstrated that these measurements are not significantly biased even if incorrect shapes and temperature distributions are assumed for the X-ray emitting regions on the stellar surface (which we hereafter call ``spots''), provided the fit of the model pulse waveform to the observed pulse waveform data is statistically acceptable. In this work, we expand on these initial studies by exploring the effect on measurements of the radius of PSR~J0740$+$6620 caused by errors in the assumed X-ray background.

The waveform of a rotating neutron star with X-ray emitting hot spots on its surface generally has both modulated and unmodulated components. In addition to counts from the hot spots, NICER also registers counts from other sources. The count rates from these other sources, such as nearby bright X-ray sources, the diffuse X-ray background from active galactic nuclei (AGN), charged particles interacting with the NICER detectors, and optical loading caused by light from the Sun, are not modulated at the neutron star's rotational frequency. The count rates produced by these other sources can therefore be treated collectively as a single, unmodulated background when analyzing pulsar observations.

Correctly determining the relative sizes of these contributions is important for the following reason. When the observed pulse waveform is analyzed, more flux not from the hot spots implies less flux from the spots. Since the modulation amplitude in the NICER data is fixed, less flux from the spots implies that the fractional modulation of the flux contributed by the spots is greater. This usually requires a larger stellar radius if the other parameters of the pulsar remain unchanged (\citealt{2016ApJ...822...27M}; \citealt{2021ApJ...918L..28M}). Hence, incorrectly estimating the flux that is not coming from the hot spots could in principle bias the radius estimate. Including in the pulse waveform analysis any information about any component of the unmodulated background should improve the accuracy of the radius estimate, as long as the information and the method used to include it are reliable \citep[see, e.g.,][]{2021ApJ...918L..26W}.

The pulsar PSR~J0740+6620 is of special interest because of its high mass ($\rm{M}\approx 2.1~\rm{M}_\odot$; see \citealt{2021ApJ...915L..12F}). The matter in its core is therefore substantially denser than the matter in the cores of $\sim 1.4~\rm{M}_\odot$ neutron stars and hence measurement of its radius provides information about the EOS of neutron star matter at densities higher than do measurements of the radii of $\sim 1.4~\rm{M}_\odot$ neutron stars. Unfortunately, the count rate measured by NICER when it is pointed at PSR~J0740$+$6620 is dominated by sources other than the hot spots on its surface, so correctly estimating the unmodulated flux contributed by these other sources is particularly important when analyzing the NICER observations of this pulsar. Important information about this unmodulated background is provided by the X-ray Multi-Mirror (XMM-Newton) blank-sky observations made over the lifetime of the mission \citep{2007A&A...464.1155C}. As discussed in \citet{2021ApJ...914L..15B,2021ApJ...918L..28M,2021ApJ...918L..27R,2022ApJ...941..150S,2024ApJ...974..294S,2024ApJ...974..295D}, knowing the XMM-Newton blank-sky background allows a more accurate estimate of the total X-ray flux from the pulsar, which helps to more accurately determine its radius. For other pulsars, this information could come from observations of the sky near the pulsar in question.

In this work, we explore the implications of assuming different unmodulated backgrounds in the XMM-Newton data, and of including this background in different ways, when estimating the radius of PSR~J0740$+$6620. To do this, we generated two sets of synthetic NICER and XMM-Newton data, one in which the XMM-Newton background is equal to the `blank-sky' background, and one in which it is much larger. We then analyzed each synthetic data set using two different pulse waveform models, one that assumes the XMM-Newton background is equal to the blank-sky background and one that allows for additional background. The construction of the synthetic data is described in Section~\ref{sec:synthetic}, and the models and our fitting procedure are described in Section~\ref{sec:models}.

We found that even if the incorrect background model is assumed in the analysis, the resulting shift of the radius posterior from the true value of the radius corresponds to only $\sim1\sigma$. In all the cases we examined, the Bayesian evidence for the model that assumes the correct XMM-Newton background is greater than the evidence for the models that assume an incorrect XMM-Newton background. This demonstrates that the Bayesian evidence can effectively discriminate between background models in otherwise ambiguous circumstances.

In Section~\ref{sec:synthetic} we describe how we generated the synthetic NICER and XMM-Newton data we analyzed, including the emission from the neutron star surface and the background. Section~\ref{sec:models} describes how we modeled the resulting synthetic data and evaluated our best-fit models, and Section~\ref{sec:Results} presents our detailed results. Our conclusions are summarized in Section~\ref{sec:conc.}.

\section{Constructing the Synthetic Data}
\label{sec:synthetic}
Our goal of this paper is to better understand the effect on the accuracy of the inferred neutron star radius of errors made in modeling  the XMM-Newton background. Because the inferred radius is obtained by jointly analyzing NICER and XMM-Newton data, we need to generate synthetic data for all the instruments that were used to collect these data. 

In constructing the two synthetic PSR~J0740$+$6620 data sets that we analyzed in this work, we assumed the same stellar rotational frequency, NICER exposure time, and XMM-Newton exposure times assumed by \citet{2021ApJ...918L..28M}.
We also assumed that the periodic amplitude modulation in the synthetic data is produced entirely by two hot spots on the stellar surface with the properties found in the analysis by \cite{2021ApJ...918L..28M}. We note, however, that emission from a pulsar's magnetosphere and shock waves around it could, in principle, also contribute to the modulation of soft X-rays from the pulsar. Although most pulsars do not exhibit such emission, the emission from the brightest, nonaccreting X-ray millisecond pulsar, PSR~J0437$-$4715, does have a significant, although faint, modulated power-law component \citep{2016MNRAS.463.2612G}. Hence, the assumption that the periodically modulated soft X-ray flux is produced only by emission from hot spots on the stellar surface is not universally valid. 

In Section~\ref{sec:synthNICER} we describe how we constructed the synthetic NICER data, in Section~\ref{sec:synthXMMblank} we explain how we generated the synthetic XMM-Newton data with a background equal to the blank-sky background, and in Section~\ref{sec:synthXMMextra} we describe how we constructed the synthetic XMM-Newton data with a background substantially larger than the blank-sky background.

\subsection{Synthetic NICER pulse waveform data}
\label{sec:synthNICER}

We created two synthetic NICER data sets independently for the two XMM-Newton backgrounds we considered. Each synthetic NICER data set was generated by Poisson sampling the pulse waveform model that \cite{2021ApJ...918L..28M} found best fit the NICER and XMM data on PSR~J0740$+$6620, when considered jointly (see Section~3.4 of \citealt{2019ApJ...887L..24M} for additional details). This model includes emission from two hot spots and an unmodulated background. The parameters in this model and their best-fit values are listed in Table~\ref{table:parameters}.

Like the actual NICER data, these two synthetic NICER data sets were two-dimensional, containing both energy and pulse-phase information. We synthesized data for NICER pulse invariant (PI) channels 30 through 123 inclusive (providing 94 energy channels) and 32 equally spaced rotational phases, for a total of 3008 phase-channel bins. 

\begin{deluxetable*}{cccc}\caption{\textbf{Primary Parameters in the Pulse Waveform Model}}
\tablehead{
\colhead{Parameter} & \colhead{Definition} & \colhead{Assumed Prior}
& \colhead{Best-Fit Value}}
\startdata
$c^2R_e/(GM)$ & Inverse stellar compactness & $3.2-8.0$ & $4.52$\\
 $M$ & Gravitational mass & $\exp[-(M - 2.08 M_\odot)^{2}/2(0.09 M_\odot)^{2}]$ & $2.067$\\
 $\theta_{c,1}$ & Spot 1 inclination of center spot & $0.0-\pi$ rad & $0.834$\\
 $\Delta\theta_1$ & Spot 1 radius & $0.0 - 3.14$ rad & $0.087$\\
 $kT_{\rm eff,1}$ & Spot 1 effective temperature & $0.011-0.5$ keV & $0.098$\\
 $\Delta\phi_2$ & Spot 1 and 2 longitudinal offset & $0.0-1.0$ cycles & $0.577$\\
 $\theta_{c,2}$ & Spot 2 inclination of center spot & $0.0-\pi$ rad & $1.223$\\
 $\Delta\theta_2$ & Spot 2 radius & $0.0-3.14$ rad & $0.066$\\
 $kT_{\rm eff,2}$ & Spot 2 effective temperature & $0.011-0.5$ keV & $0.103$\\
 $\theta_{\rm obs}$ & Observer inclination & $1.44-1.62$ rad & $1.561$\\
 $N_H$ & Column density & $0.0-2.0 \times 10^{20}$ cm$^{-2}$ & $0.006$\\
 $D$ & Distance & $\exp[-(D - 1.136 \rm{kpc})^{2}/2(0.20 \rm{kpc})^{2}], D \geq 1.136 \rm{kpc}$ & $1.151$\\ 
  & & $\exp[-(D - 1.136 \rm{kpc})^{2}/2(0.18 \rm{kpc})^{2}], D \leq 1.136 \rm{kpc}$ & 
\enddata
\tablecomments{This table details the pulse waveform model that \citet{2021ApJ...918L..28M} fit to the NICER and XMM-Newton data on PSR~J0740$+$6620, the priors on the parameters (with the exception that \citealt{2021ApJ...918L..28M} used flat priors of 0.0--3.0~rad for $\Delta\theta_1$ and $\Delta\theta_2$), and the values of the parameters in this model that they found gave the best joint fit to the NICER and XMM-Newton data.  The 12 parameters listed here are the primary parameters in the pulse waveform model used in \citet{2021ApJ...918L..28M} and in the present work; $R_e$ is the equatorial circumferential radius.  We generated synthetic pulse waveform data by sampling this best-fit waveform, assuming the same 346.532~Hz rotational frequency assumed by \citet{2021ApJ...918L..28M}.   The priors listed here are the same as those we used when fitting this model to the two synthetic data sets we analyzed. See the text for further details.}
\label{table:parameters}
\end{deluxetable*}

\subsection{Dataset 1: Synthetic XMM-Newton data includes only the blank-sky background}
\label{sec:synthXMMblank}

We used the XMM-Newton ``blank-sky'' background to estimate the background present in the XMM-Newton data on PSR~J0740$+$6620 because the relatively short durations of the observations of PSR~J0740$+$6620 by the XMM-Newton instruments (roughly 18.0~ks for MOS1, 18.7~ks for MOS2, and 6.8~ks for pn) yielded too few background counts to accurately estimate the background. XMM-Newton's low instrumental background, high angular resolution, and wide field of view make it a good instrument for making blank-sky observations. The XMM-Newton ``blank-sky'' background therefore provides a reliable estimate of the blank-sky component of the background in XMM-Newton observations of PSR~J0740$+$6620.
 
The XMM-Newton ``blank-sky'' background has been constructed using data obtained by observing apparently empty regions of the sky over $\sim 20$~yr. These observations totaled $\sim$~1.5 Ms for the two European Photon Imaging Camera Metal-Oxide-Silicon detectors (MOS1 and MOS2) and $\sim 450$~ks for the pn CCD (pn) \citep{2007A&A...464.1155C}. These mission-averaged data files provide a standard estimate of the detector and diffuse X-ray backgrounds. In constructing our synthetic data, we filtered the event files from the XMM-Newton blank-sky observations identically to the way we filtered the event files from the XMM-Newton observations of PSR~J0740$+$6620. We then extracted the background counts and rescaled the number of counts to match the PSR~J0740$+$6620 exposure times.

One XMM-Newton synthetic data set we investigated assumed that the only XMM-Newton counts that did not come from the hot spots are ``blank-sky'' counts. The full synthetic XMM-Newton data for this case were generated by Poisson-sampling the sum of our best-fit hot-spot counts and blank-sky background for each of the MOS1, MOS2, and pn detectors. For the MOS1 and MOS2 detectors, we used PI channels 20 through 99 inclusive, while for the pn detector we used PI channels 57 through 299 inclusive, as in \citet{2021ApJ...918L..28M}. As noted previously, the XMM-Newton data do not have sufficient time resolution for these counts to be assigned to the finely spaced phase bins we used to analyze the NICER data. Hence, no phases were assigned to the XMM-Newton counts. Therefore, a model in which all the XMM-Newton counts are attributed to an unmodulated background with none attributed to the hot spots could, in principle, fit these XMM-Newton data perfectly. This emphasizes the importance of determining the actual background.

\subsection{Dataset 2: Synthetic XMM-Newton data includes additional background}
\label{sec:synthXMMextra}

We also investigated a second XMM-Newton synthetic data set that assumed the number of XMM-Newton counts that did not come from the hot spots is substantially greater than the ``blank-sky'' counts. We constructed this second XMM-Newton synthetic data set by adding additional background counts to the synthetic blank-sky counts in the MOS1, MOS2, and pn detectors. We generated these additional counts by applying a first-order Savitzky-Golay filter \citep{1964AnaCh..36.1627S} to the \emph{total} counts collected by these detectors when XMM-Newton was pointed at PSR~J0740$+$6620. We used filters with window lengths of 130 channels for the pn data and 50 channels for the MOS1/MOS2 data. We then performed a Poisson draw from this fit and added the resulting counts, channel by channel, to the synthetic data. 

Comparing the resulting number of counts in each energy channel of the three XMM-Newton detectors with the number of counts in each channel shown in Figure~8 of \citet{2021ApJ...918L..28M} for their model of XMM-Newton's ``blank-sky'' counts, the average number of XMM-Newton background counts is a factor of $\sim 5$ greater. As a result, the total number of counts in this synthetic XMM-Newton background data is, on average, twice the total number of counts XMM-Newton detected when it was pointed in the direction of PSR~J0740$+$6620. We chose this synthetic XMM-Newton background data to explore the effect on the estimate of the radius of PSR~J0740$+$6620 if the XMM-Newton background is substantially greater than is normally estimated. Such a larger background could be due, for example, to cosmic rays, X-ray bright stars, or AGN. 

\section{Modeling the Synthetic Data}
\label{sec:models}

Following \citet{2021ApJ...918L..28M} and \citet{2024ApJ...974..295D}, we modeled the pulse waveform in our synthetic data using two uniform-temperature circular spots and assuming a nonmagnetic hydrogen atmosphere. For the analyses reported here, we used the atmospheric tables for  emission by partially ionized hydrogen from \citet{2009Natur.462...71H} (see \citealt{2005MNRAS.360..458B} for the relevant opacity tables).  This pulse waveform model has 12 primary parameters, which specify the mass and compactness of the neutron star, the radii, temperatures, and locations of the two spots, the phase offset between them, the observer's inclination, and the column density and distance to the pulsar. Table~\ref{table:parameters} lists these 12 parameters, the ranges of the priors \citet{2021ApJ...918L..28M} used when they fit this pulse waveform model to the actual NICER and XMM-Newton data, and their best-fit values of these parameters. As Table~\ref{table:parameters} notes, when we fit this model to the synthetic data considered in the present study, we assumed slightly wider priors for $\Delta\theta_1$ and $\Delta\theta_2$.

Although the XMM-Newton data are unmodulated, these data are important in constraining the phase-independent background in each energy channel. This constraint helps disentangle astrophysical backgrounds from nonmodulated surface emission, yielding more precise constraints on the hot-spot properties and stellar compactness. In particular, knowing the background level from XMM-Newton allows the model to attribute the correct fraction of the observed counts in each channel to the modulated emission from the hot spots, tightening the posteriors on their temperatures and sizes and thereby tightening the constraints on mass--radius combinations.

In fitting this model to the actual NICER and XMM-Newton data on PSR~J0740$+$6620, \citet{2021ApJ...918L..28M} labeled the two spots `1' and `2' and allowed them to overlap without restriction. To avoid ambiguity when a given pixel is covered by both spots, they assumed that such a pixel emits with the effective temperature of spot 1. They also assumed a stellar rotational frequency of 346.532 Hz, a NICER exposure time of $1,602,683.7$ s, an XMM-pn exposure time of $6,808.74$ s, an XMM-MOS1 exposure time of $17,959.5$ s, and an XMM-MOS2 exposure time of $18,680.7$ s \citep{2021ApJ...918L..28M,2021ApJ...918L..27R}. We followed this same procedure when fitting the two waveform models considered in the present work to the two synthetic data sets we generated.

\begin{figure}[htbp]
\centering
\vspace{-13pt}
\gridline{\fig{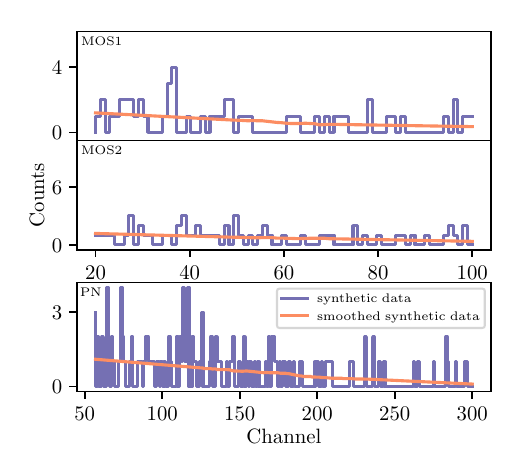}{\columnwidth}{(a)}}
\vspace{-14pt}
\gridline{\fig{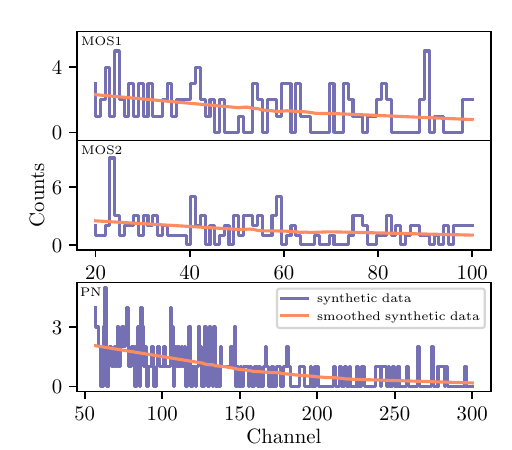}{\columnwidth}{(b)}}
\vspace{-8 pt}
\caption{The data we used to define the priors on the backgrounds when fitting synthetic Dataset 1 and synthetic Dataset 2. In panel (a) the purple histograms show the counts that were generated by Poisson sampling the sum of the modeled spot emission and the XMM-Newton blank-sky background, while the orange curves show the same data smoothed using a Savitzky--Golay filter. When we analyzed synthetic Dataset 1, which includes only the XMM-Newton blank-sky background, using a model that allows for an additional unmodulated background, we used the smoothed data to define the prior for that additional background. In panel (b) the purple histograms show the counts generated by Poisson sampling the sum of the modeled spot emission, the XMM-Newton blank-sky background, and the additional unmodulated background we considered. Again, the orange curves show the smoothed data. In this case, we used the smoothed data to define the priors on the additional backgrounds when we analyzed synthetic Dataset 2 using a model that includes the XMM-Newton blank-sky background and an additional unmodulated background.}
\label{fig:smooth}
\end{figure}

The two full pulse waveform models we fit to synthetic Dataset 1 and Dataset 2 each contain the same 12 primary waveform parameters (see Table~\ref{table:parameters}), but differ in the assumptions we made about the phase-independent background in the XMM-Newton data. In one model, we assumed that the only non-spot contribution is the XMM-Newton blank-sky background, with the background in each energy channel modeled as a Poisson-distributed variable whose prior is given by the long-term blank-sky measurements. In the other model, we allowed for additional unmodulated background in each energy channel, in addition to the blank-sky counts. In both models, the NICER background is treated identically: we marginalized over a phase-independent term in each energy channel using a Gaussian fit to the likelihood profile (see Section~3.4 of \citealt{2019ApJ...887L..24M} for additional details). Figure~\ref{fig:smooth} shows the smoothed synthetic spectra we used to define the priors on the additional XMM-Newton background for the two synthetic data sets.

Having specified the two models and their priors, we then fit each model to each synthetic data set. For each combination of model and data set, we performed a maximum likelihood analysis, determining the values of the 12 primary waveform parameters and the background parameters that maximized the total log likelihood given the data. The background parameters affect the predicted counts in each energy channel and thus the log-likelihood contributions from each instrument. For example, increasing an assumed background reduces the contribution from the hot spot, altering the modulation amplitude and shifting the maximum of the likelihood to a larger radius. 

\subsection{Calculation of the XMM-Newton log likelihood when the only background is assumed to Be from the blank sky}
\label{sec:backonlyblank}

For our first analysis, we assumed that all the counts that do not come from the two hot spots are background counts and that these counts are fully described by the counts collected by XMM-Newton during its blank-sky observations.  This was the assumption made by \citet{2021ApJ...918L..28M} and \citet{2024ApJ...974..295D} when they analyzed the NICER and XMM-Newton data on PSR~J0740$+$6620. We treat the XMM-Newton blank-sky data as a Poisson-constrained prior on the background, scaled to the observation and marginalized over in the likelihood, following the procedure used in \citet{2021ApJ...918L..28M}. This approach assumes that the background is statistically stationary over the relevant timescales, such that the blank-sky data provide a representative realization of the background during the observation.

XMM-Newton has made blank-sky observations for more than two decades, but the total number of counts it has collected in each energy channel of the XMM-pn, XMM-MOS1, and XMM-MOS2 detectors during these observations is small, so it is necessary to use Poisson statistics when treating them. The probability that the long-term average number of blank-sky counts over a time $T_{\rm bk}$ is between $m_{\rm bk}$ and $m_{\rm bk}+dm_{\rm bk}$ is
\begin{equation}
{\rm prob}(m_{\rm bk})dm_{\rm bk}=\frac{m_{\rm bk}^{N_{\rm bk}}}{N_{\rm bk}!}e^{-m_{\rm bk}}dm_{\rm bk}\; .
\end{equation} 
For our synthetic Dataset 1, we adopted this probability as the prior for the blank-sky counts in this energy channel and used it to estimate the number of blank-sky counts that, in addition to the counts from the hot spots, produced the total number of counts in that energy channel. Then, for an observation lasting a time $T_{\rm obs}$, the expected number of blank-sky background counts is $b_{\rm bk}=m_{\rm bk}(T_{\rm obs}/T_{\rm bk})$.

If the number of counts from the hot spots in the specified detector channel is $s$, then the total number of counts predicted by the model (including both the counts contributed by the spots and the counts contributed by this candidate background) is $s+b_{\rm bk}$. For a detector channel with $N_{\rm obs}$ counts observed over time $T_{\rm obs}$, the Poisson likelihood of the data, given these hot spot and background models, is 
\begin{equation}
{\cal L}(N_{\rm obs}|s+b_{\rm bk})=\frac{(s+b_{\rm bk})^{N_{\rm obs}}}{N_{\rm obs}!}e^{-(s+b_{\rm bk})}\; .
\end{equation}
The posterior probability is then proportional to the product of the probability of $m_{\rm bk}$ and the likelihood:
\begin{equation}
\begin{array}{rl}
P(s+b_{\rm bk}|N_{\rm obs},N_{\rm bk})&\propto \frac{m_{\rm bk}^{N_{\rm bk}}}{N_{\rm bk}!}e^{-m_{\rm bk}}\\ \\
&\times \frac{(s+b_{\rm bk})^{N_{\rm obs}}}{N_{\rm obs}!}e^{-(s+b_{\rm bk})}\;.
\label{eq:posteriorblank}
\end{array}
\end{equation}
The likelihood of the data given the model is then given by marginalizing $P(s+b_{\rm bk}|N_{\rm obs},N_{\rm bk})$ over $m_{\rm bk}$.

\subsection{Calculation of the XMM-Newton log likelihood when We allow for the possibility of additional background}
\label{sec:backextra}

For our second analysis, we incorporated the possibility of additional unmodulated background in the XMM-Newton data on PSR~J0740$+$6620. In addition to the XMM-Newton blank-sky background, we included an extra component that could account for sources such as backgrounds from nearby sources or intrabinary shocks in the pulsar system. The maximum allowed number of counts $b_{\rm max}$ from this additional background, in any given energy channel for each instrument, was taken to be the smoothed count rate predicted from the pointed XMM-Newton observation in that channel. This smoothed spectrum was obtained by applying a Savitzky--Golay filter \citep{1964AnaCh..36.1627S} to the total counts in the synthetic data. This procedure produces an estimate of the mean counts per channel that serves as a physically motivated upper bound for the additional background, ensuring that the inferred additional background does not exceed the average counts measured in the observation.

We assumed that the prior on the additional background $b_{\rm other}$ in each channel was uniform between zero and the corresponding smoothed value $b_{\rm max}$:
\begin{equation}
{\rm prob}(b_{\rm other})db_{\rm other}=\frac{1}{b_{\rm max}}db_{\rm other}\; .
\end{equation}
This choice is independent of the specific hot-spot waveform model and is therefore acceptable for model comparison. When a model includes this additional background, the predicted counts per detector energy channel are given by the Poisson-sampled $s+b_{\rm bk}+b_{\rm other}$. The posterior probability for the model parameters in this case is then
\begin{align}\nonumber
P(s+b_{\rm bk}+b_{\rm other}|N_{\rm obs},N_{\rm bk})\propto\\
\frac{m_{\rm bk}^{N_{\rm bk}}}{N_{\rm bk}!}e^{-m_{\rm bk}}
\frac{1}{b_{\rm max}}e^{-(s+b_{\rm bk}+b_{\rm other})} \nonumber
\\ \times \frac{(s+b_{\rm bk}+b_{\rm other})^{N_{\rm obs}}}{N_{\rm obs}!} .
\label{eq:posteriorother}
\end{align}
We then obtained the likelihood of the data given the model by marginalizing  $P(s+b_{\rm bk}+b_{\rm other}|N_{\rm obs},N_{\rm bk})$ over both $m_{\rm bk}$ and $b_{\rm other}$. This approach provides a way to test how the presence of such a background, particularly if it is not included in the model, can influence the inferred radius.

\subsection{Fitting our two different pulse waveform models to Datasets 1 and 2}\label{sec:fitting}

We fit each of our two different full pulse waveform models to our two different full synthetic data sets (Data Set~1 and Data Set~2). Specifically, for each full pulse waveform model and synthetic data set, we calculated the log Poisson likelihood of the model by summing the log-likelihoods for each of the four data sets. The total log-likelihood is then
\begin{align}
\nonumber    \log{\cal L}_{\rm{tot}}=&\log{\cal L}_{\rm{NICER}}+\log{\cal L}_{\rm{XMM_{PN}}}\\&+\log{\cal L}_{\rm{XMM_{MOS1}}}+\log{\cal L}_{\rm{XMM_{MOS2}}}\; .
    \label{eq:logl}
\end{align}
Equation~(\ref{eq:logl}) assumes that the four different data sets are independent, which is reasonable given that NICER and XMM-Newton are separate instruments and that the XMM-MOS1, XMM-MOS2, and XMM-pn detectors operate independently, using distinct hardware and readout systems. For each data set, we computed the log likelihood by summing the log Poisson likelihood across all the energy channels and (for NICER) all the rotational phase bins (see Section 3.4 in \citealt{2021ApJ...918L..28M}).

We determined the pulse waveform models that best fit the joint NICER and XMM-Newton synthetic data by adjusting the 12 primary pulse waveform parameters listed in Table~1 to maximize the log posterior.  In addition to the primary pulse waveform parameters, our fitting procedure introduced additional nuisance parameters.  

For the part of the fit  that was to the XMM-Newton synthetic data, these nuisance parameters describe the background, and are discussed in Section~\ref{sec:backonlyblank} (when the only non-spot contribution is the blank-sky background) and Section~\ref{sec:backextra} (when we allow for the possibility of additional background).  

For the part of the fit that was to the NICER data, the rotational phase introduces an additional nuisance parameter and changes our approach to marginalizing over the background. This additional nuisance parameter is the overall rotational phase of the pulse waveform. To determine this phase, we computed a trial pulse waveform and then determined the log-likelihood of the data given the model for different assumed rotational phases. We then fitted a Gaussian to the resulting likelihood distribution and marginalized over the phase.  

For the NICER background, we did not impose prior constraints on the unmodulated backgrounds present in the NICER data, although models for these backgrounds exist (e.g., the 3C50 model of \citealt{2022AJ....163..130R}). Instead, we included a phase-independent background independently for each energy channel, fitted a Gaussian to the likelihood as a function of the added backgrounds, and then marginalized.

Our models involve a substantial number of parameters, making posterior exploration in the resulting high-dimensional spaces a significant computational challenge. To address this challenge, we adopted a hybrid sampling approach.

We used the nested sampling algorithm \texttt{MultiNest} \citep{2008MNRAS.384..449F, 2009MNRAS.398.1601F, 2016ascl.soft06005B} but only as a starting point for our posterior sampling. Previous studies \citep[e.g.,][]{2019ApJ...887L..24M, 2021ApJ...918L..28M, 2021AJ....162..237I, 2023MNRAS.521.1184L, 2024OJAp....7E..79D, 2024ApJ...974..295D} have found that \texttt{MultiNest} often produces posterior credible regions that are too narrow, therefore underestimating the uncertainties in the parameters, when the number of live points $N_{\rm live}$ is insufficient and/or the sampling efficiency parameter $\eta$ is too large \citep[see, e.g.,][]{2024ApJ...974..295D}. 

As an example, \citet{2024ApJ...974..294S} reported a headline radius of $12.49^{+1.28}_{-0.88}$~km using $N_{\rm live} = 4 \times 10^4$ and $\eta = 0.01$ when they required the radius to be less than 16~km, but in an alternative run using a smaller sampling efficiency ($\eta = 10^{-4}$) and fewer live points, the radius uncertainty broadened to $12.55^{+1.37}_{-0.92}$~km, although they did not carry out analyses to the point of achieving converged and robust uncertainties. \citet{2024ApJ...974..295D}, who used the same nested sampling + Markov Chain Monte Carlo (MCMC) strategy as our current investigation, found $12.76^{+1.49}_{-1.02}$~km when they imposed the same ad hoc radius limits as \citet{2024ApJ...974..294S}, though the full range was $12.92^{+2.09}_{-1.13}$~km. We also note that \citealt{2025PhRvD.112b3008H} provide a clear illustration of the region-sampling limitation of \texttt{MultiNest} in their Figure 1 (left panel), which shows how the multi-ellipsoidal approximation can miss portions of the likelihood surface.

For our initial analysis of the pulse waveforms studied here, we used $\eta = 0.01$ and $N_{\rm live} = 4096$, following \citet{2024ApJ...974..295D}. While these choices for the values of these parameters made our initial exploration of the parameter space efficient, they did not necessarily lead to robust posteriors. As an example, for our initial analysis we carried out three independent \texttt{MultiNest} runs. The injected values of only 18 of the 36 parameters (6 of 12 per realization) fell within the $\pm 1\sigma$ credible region predicted by the \texttt{MultiNest}-only analyses, versus the $\approx 24$ expected given that the $\pm 1\sigma$ band contains 68\% of the probability. This is consistent with the previously observed tendency for \texttt{MultiNest} to underestimate posterior uncertainties.

To refine our results, we initialized parallel tempered MCMC (\texttt{ptemcee}; \citealt{2013PASP..125..306F}) using walkers from the completed \texttt{MultiNest} run, thereby leveraging its output to accelerate convergence of the \texttt{ptemcee} analysis. \texttt{ptemcee} ensures detailed balance, allowing the MCMC chains to asymptotically sample the true posterior distribution. We assessed convergence by monitoring the 1st, 16th, 50th, 84th, and 99th percentiles of each posterior.

Our settings included 1024 walkers per temperature rung across four rungs, with temperatures equal to 1 (which thus produces an unmodified log likelihood), 2, 4, and 8 (totaling $1024\times4 = 4096$ walkers). This approach allowed us to combine the rapid exploration of the parameter space that \texttt{MultiNest} allows with the convergence properties of \texttt{ptemcee}. We verified robustness by repeating the combined \texttt{MN}+\texttt{ptemcee} analysis three times independently, each yielding consistent posteriors. After \texttt{ptemcee} continuation, 21 of the 36 injected values across the three independent runs fall within their $1\sigma$ credible regions, compared to 18 of 36 from \texttt{MultiNest} alone and closer to the expected $\sim$24 of 36. We note that the reliability of \texttt{MultiNest} depends sensitively on the shape of the likelihood surface, and that adequate parameter recovery for the aggregate of all the model parameters does not guarantee reliable inference for individual parameters with complex posterior surfaces, such as the radius and spot colatitudes.

\begin{deluxetable}{cccc}
\tablewidth{\linewidth}
\caption{\textbf{Widths of the Radius Distributions from \texttt{MultiNest}-only and \texttt{MultiNest}+\texttt{ptemcee} Analyses}}
\tablehead{
\colhead{Case} & \colhead{$\rm{\Delta R_{e,MN}}$ (km)} & \colhead{$\rm{\Delta R_{e,MN+PT}}$ (km)} & \colhead{\% Broadened}}
\startdata
1 & 3.23 & 4.59 & 42.11 \\
2 & 2.85 & 3.18 & 11.58 \\
3 & 4.41 & 5.89 & 33.56 \\
4 & 4.71 & 5.98 & 26.96 \\
\enddata
\tablecomments{Comparison of the $\pm 1 \sigma$ credible region width ($\rm{\Delta R_e}$), defined as the width of the $\pm 1 \sigma$ credible region for the radius, between MultiNest-only ($\rm{\Delta R_{e,MN}}$) and \texttt{MultiNest}+\texttt{ptemcee} ($\rm{\Delta R_{e,MN+PT}}$) analyses.
See~Table~\ref{table:totresultsR} for the definitions of the cases listed here.
}
\label{table:width}
\end{deluxetable}

\begin{deluxetable*}{ccccccccc}
\setlength{\tabcolsep}{10pt}
\tablewidth{\linewidth}
\caption{\textbf{Summary of Equatorial Radius Posteriors from \texttt{MultiNest}+\texttt{ptemcee} Analyses}}
\tablehead{
\colhead{Case} & \colhead{Data} & \colhead{Model} & \colhead{Input $R_e$} & \colhead{$-2\sigma$ $R_e$} & \colhead{$-1\sigma$ $R_e$} & \colhead{Median $R_e$} & \colhead{$+1\sigma$ $R_e$} & \colhead{$+2\sigma$ $R_e$}}
\startdata
1 & Extra\tablenotemark{a}  & Extra\tablenotemark{b} & 13.823 & 11.29 & 12.7 & 14.61 & 17.29 & 20.76 \\
2 & Extra\tablenotemark{a} & Standard\tablenotemark{d} & 13.823 & 9.67 & 10.38 & 11.5 & 13.56 & 17.25 \\ 
3 & Standard\tablenotemark{c} & Extra\tablenotemark{b} & 13.823 & 12.44 & 14.69 & 17.3 & 20.58 & 23.51 \\
4 & Standard\tablenotemark{c} & Standard\tablenotemark{d} & 13.823 & 10.84 & 12.32 & 14.7 & 18.3 & 22.3 \\
\enddata
\tablenotemark{a}{See Section \ref{sec:synthXMMextra}}
\tablenotemark{b}{See Section \ref{sec:backextra}}
\tablenotemark{c}{See Section \ref{sec:synthXMMblank}}
\tablenotemark{d}{See Section \ref{sec:backonlyblank}}
\tablecomments{All equatorial radius values are given in km. We generated one synthetic data set for each background model (Datasets 1 and 2) using the best-fit model of the PSR~J0740$+$6620 pulse waveform described in the text and the specified background model. The pulse waveform model assumes two circular spots in a nonmagnetic, partially ionized hydrogen atmosphere. In Cases 1 and 4, the synthetic data were analyzed assuming the same background model that was used to generate the synthetic data, whereas in Cases 2 and 3 the synthetic data were analyzed assuming an incorrect background model (see the text for details). Inferred equatorial circumferential radii (in kilometers) are those given by the fits listed in columns (2) and (3). Definitions: Dataset 1 = XMM-Newton blank-sky background only; Dataset 2 = XMM-Newton blank-sky background plus an additional unmodulated background. Background Model 1 = blank-sky background only; Background Model 2 = blank-sky background plus an additional unmodulated background. Cases 1–4 correspond to the four possible combinations of \rm{Dataset} and \rm{Background Model}. The radius assumed in the model used to generate the synthetic pulse waveform data is within 1 standard deviation of the median inferred radius for Cases 1 and 4, and just outside one standard deviation for Cases 2 and 3.}
\label{table:totresultsR}
\end{deluxetable*}

To assess our approach, we compared the width of the $\pm 1\sigma$ credible regions for the inferred radius, which we denote as $\rm{\Delta R_e}$, given by the \texttt{MultiNest}-only and the \texttt{MultiNest}+\texttt{ptemcee} analyses for the four cases we investigated. Table~\ref{table:width} summarizes the results; see~Table~\ref{table:totresultsR} for the definitions of these cases. In every case, the initial, MultiNest-only estimate of $\rm{\Delta R_e}$ was appreciably broadened by subsequent use of \texttt{ptemcee}, reinforcing previous findings that \texttt{MultiNest} tends to give erroneously narrow credible intervals when $N_{\rm live}$ is too small and/or $\eta$ is too large.

\subsection{Evaluating the models}\label{sec:evalmodels}

To evaluate our models, we used $\chi^2$ tests to assess the absolute goodness of the fit and the Bayesian evidence to compare models.

We computed the chi-squared statistic $\chi^2$ using the model variance form originally proposed by \citet{FRS2009XOT}, namely,
\begin{equation}
\chi^2 = \sum_{i} \frac{(m_i - d_i)^2}{m_i}.
\end{equation}
Here, $m_i$ is the expected number of counts in bin $i$ predicted by the model, and $d_i$ is the observed number of counts in bin $i$. This formulation differs from the data variance form, which replaces $m_i$ in the denominator with $d_i$. Unlike the data variance form, the model variance form ensures that when the model is correct and $d_i$ follows a Poisson distribution with mean $m_i$, the expected value of $\chi^2$ equals the number of bins, regardless of $m_i$, although when $m_i$ is small the $\chi^2$ distribution is broader than when $m_i$ is large. 

Our assessments of the implications of the value of $\chi^2$ apply only to the fits of the pulse waveform model to the NICER data---the XMM-Newton data have too few counts for an assessment of $\chi^2$ to be meaningful.  We compute $\chi^2$ for the NICER part of the fit using both the full phase-channel data and the bolometric data; because $\chi^2$ is a nonlinear measure, these provide different perspectives on the goodness of fit.  For example, a fit which is consistently slightly high at a given phase will have much larger bolometric $\chi^2$ at that phase than the sum of all of the channel-phase $\chi^2$ values at that phase.

The lower the probability of $\chi^2$ for the number of degrees of freedom, the poorer the fit. We took a probability less than $10^{-3}$ as indicating that the model may be inadequate. Conversely, a probability in the tens of percent range does not confirm correctness but suggests consistency between the model and the data. Intermediate probabilities require careful consideration in the context of the analysis.

The Bayesian evidence ($Z$) serves a different purpose by directly comparing models while accounting for model goodness of fit and complexity. Its logarithmic form, $\log Z$, is defined as
\begin{equation}
\log Z = \log \int \mathcal{L}(\theta) \pi(\theta) \, d\theta,
\end{equation}
where $\mathcal{L}(\theta)$ is the likelihood function, $\pi(\theta)$ is the normalized prior probability distribution of the parameters $\theta$, and the integral spans the parameter space. Since direct evaluation of this integral is often intractable, numerical techniques such as nested sampling and parallel tempering can help obtain useful results. Nested sampling estimates $\log Z$ by iteratively replacing low-likelihood samples with new ones at increasing likelihood thresholds, effectively transforming the integral into a weighted sum. Parallel tempering runs Markov chains at different temperatures, facilitating efficient exploration of the parameter space and improving the accuracy of the integration. A larger value of $\log Z$ indicates stronger support for a given model relative to alternatives \citep[e.g.,][]{sivia2006data}.

Because the Bayesian evidence inherently involves comparing models, it complements $\chi^2$, which measures how well a single model fits the data on its own, without directly evaluating alternative models. However, the value of the evidence can be sensitive to the prior volumes adopted for all the model parameters, since it is computed by integrating over the full prior space. Assuming broader priors for the parameters increases the volume of the prior space, which can penalize the evidence, even if the model fits the data well. In the present analysis, the parameters that affect the volume of the prior space include the background parameters discussed earlier. To ensure that model comparisons based on $\log Z$ are robust, it is important to check that any preference for one model over another persists across a range of reasonable choices for the priors of their parameters.

\section{Results}
\label{sec:Results}

We now present the results of our analyses of the four cases listed in Table~\ref{table:totresultsR}. For each case, we list the assumption about the background that was made when the synthetic data was generated and the assumption about the background that was made in the model that was used to analyze the synthetic data. Each case explores how the relationship between the background in the synthetic data and the background assumed in the fitted model affects the estimated radius of the star. These results improve our understanding of when and how a mismatch between the actual (in this case, synthetic) background and the background assumed when analyzing the data affects the inferred stellar radius.

Table~\ref{table:totresultsR} reports the inferred median values and $\pm 1\sigma$ and $\pm 2\sigma$ credible regions for the equatorial circumferential radii given by these analyses. Table~\ref{table:results} shows the quality of the fits given by the phase-channel and bolometric $\chi^2$ values and compares the Bayesian evidence for these fits computed using \texttt{MultiNest}.  To verify the values of the Bayesian evidence computed using \texttt{MultiNest} we independently computed the Bayesian evidence using \texttt{pocoMC} \citep{2022JOSS....7.4634K}. In all cases, the logarithmic evidence computed using \texttt{MultiNest} and \texttt{pocoMC} differed by less than 0.7, confirming the consistency of the evidence values computed using these two codes. In the following subsections, we describe our key results for each of the four cases listed in Table~\ref{table:totresultsR}.

\begin{figure*}
\begin{subfigure}{.491\textwidth}
      \includegraphics[width=\linewidth]{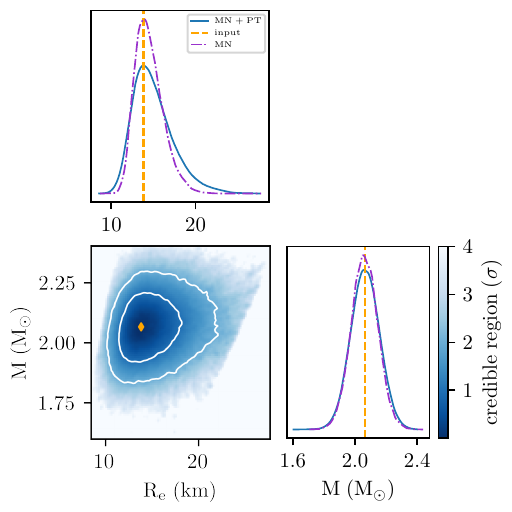}
      \caption{Case 1: The synthetic data were generated with a background approximately five times the XMM-Newton blank-sky background, and the fitted model allowed for an additional background beyond the blank-sky component. The true radius lies at the 36th percentile of the \texttt{ptemcee} radius posterior, near the peak of the distribution.}
      \label{fig:main1}
\end{subfigure}
\hspace{0.01\textwidth}
\begin{subfigure}{.491\textwidth}
      \includegraphics[width=\linewidth]{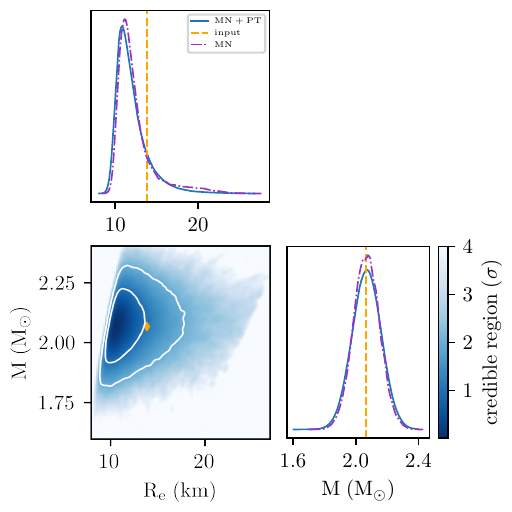}
     \caption{Case 2: The synthetic data were generated with a background approximately five times the XMM-Newton blank-sky background, but the fitted model assumed only the blank-sky background. The radius posterior peaks at a value smaller than the true radius used to generate the synthetic data, which lies at the 86th percentile of the posterior distribution.}
      \label{fig:main2}
\end{subfigure}
\caption{The mass and radius posterior probability distributions from our combined \texttt{MultiNest} and \texttt{ptemcee} analysis for Case 1 (a) and Case 2 (b). In each case, the orange diamond in the 2D plot and the orange dashed lines in the 1D plots mark the mass and radius values assumed in generating the synthetic data. Solid blue and dot--dashed purple curves show the \texttt{ptemcee} and \texttt{MultiNest} posteriors, respectively.  As discussed in the text, the radius distribution broadens from the initial \texttt{MultiNest} result to the converged \texttt{ptemcee} distribution, while the mass posterior remains dominated by the prior.}
\end{figure*}

\begin{deluxetable}{cccc}
\setlength{\tabcolsep}{2pt}
\tablewidth{\linewidth}
\caption{\textbf{Quality of Fits to the Synthetic Data}}
\tablehead{
\colhead{Case} & \colhead{$(\chi^2/{\rm dof})_{PE}$ (Prob.)} &  \colhead{$(\chi^2/{\rm dof})_B$ (Prob.)} & \colhead{$\Delta \rm logZ$}}
\startdata
1 & 2978.47/2901 (0.155) & 29.85/25 (0.23) & - \\
2 & 2980.24/2901 (0.149) & 29.01/25 (0.26) & -8.59 $\!\pm\!$ 0.11 \\ 
\hline
3 & 2920.28/2901 (0.397) & 39.24/25 (0.035) & -5.76 $\!\pm\!$ 0.10 \\
4 & 2916.59/2901 (0.416) & 37.41/25 (0.053) & - \\
\enddata
\tablecomments{$\chi^2$ values for the phase- and energy-resolved (PE) and bolometric (B) pulse waveforms and the probabilities of the best-fit models for each case. Fitting the models that were used to generate the synthetic data (Cases 1 and 4) yielded higher log evidences than fitting the mismatched model-data combinations (Cases 2 and 3); hence, the former models are preferred. This suggests that a converged analysis can be a useful tool to identify the more promising models. All the fits to the bolometric and phase-energy channel pulse waveform appear acceptable. Note that these $\chi^2$ assessments apply only to the fits of the pulse waveform model to the NICER data. None of the fits, even the ones that assumed incorrect backgrounds, provide a clear mismatch with the data. Note that the $\chi^2$ values reported here are evaluated at the maximum likelihood parameter vector from the NICER and XMM-Newton fit, including marginalization over the XMM-Newton background; they are not the minimum possible NICER $\chi^2$ values, and it is therefore possible for a more flexible model (e.g., Case 3) to yield a larger $\chi^2$ than a more restricted model (e.g., Case 4) at their respective maximum likelihood solutions.}
\label{table:results}
\end{deluxetable}

\subsection{The Synthetic Data Include and the Fitted Model Allow an Additional Background}\label{subsec:1}

In Case 1, the synthetic XMM-Newton data were generated with a background larger than the standard XMM-Newton blank-sky background, and the fitted model explicitly allowed for this additional background. As expected for a model matched to the data generation process, the inferred neutron star radius is accurate (see Table~\ref{table:results} and Figure~\ref{fig:main1}), and both the phase-channel and bolometric $\chi^2$ values indicate satisfactory fits.

The phase-channel $\chi^2$ quantifies the agreement between the model and the data across the full two-dimensional grid of rotational phase and energy, whereas the bolometric $\chi^2$ provides a complementary measure by integrating over energy and comparing the phase-resolved pulse profiles. For the phase-energy-resolved fit, the best-fitting two-spot model yielded $\chi^2 = 2978.47$ for 3008 phase–energy bins. Accounting for 12 primary parameters, 94 background-related nuisance parameters, and one overall phase parameter leaves $3008 - 94 - 12 - 1 = 2901$ degrees of freedom, giving $\chi^2/dof = 2978.47/2901$. If the model is correct, the probability of obtaining a $\chi^2$ at least this large is $p = 0.155$, which exceeds our threshold for considering a fit acceptable (Section \ref{sec:evalmodels}).

\begin{figure*}
\begin{subfigure}{.491\textwidth}
      \includegraphics[width=\linewidth]{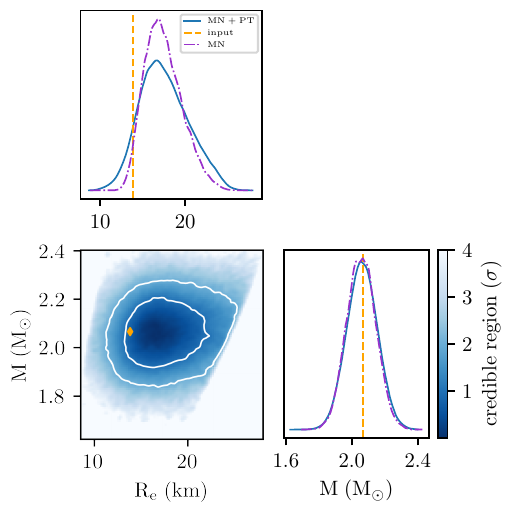}
      \caption{Case 3: In this case the synthetic data contained only the XMM-Newton blank-sky background, whereas the fitted model allowed for an additional unmodulated background. This assumption leads to a posterior distribution that peaks at a radius larger than the true input value, with the true radius lying at the 8th percentile of the posterior distribution. }
      \label{fig:main3}
\end{subfigure}
\hspace{0.01\textwidth}
\begin{subfigure}{.491\textwidth}
      \includegraphics[width=\linewidth]{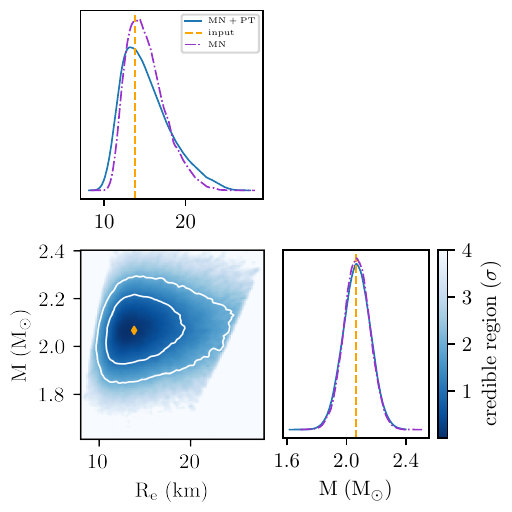}
      \caption{Case 4: We generated the synthetic XMM-Newton data assuming a background comparable to the blank-sky background, and also assumed this when we fit the synthetic data. The posterior peaks around the radius value assumed during synthetic data generation, with the true radius lying at the 38th percentile of the posterior distribution.}
      \label{fig:main4}
\end{subfigure}
\caption{The mass and radius posterior probability distributions from our combined \texttt{MultiNest} and \texttt{ptemcee} analysis for Case 3 (a) and Case 4 (b). In each case, the orange diamond in the 2D plot and the orange dashed lines in the 1D plots mark the mass and radius values assumed in generating the synthetic data.}
\end{figure*}

The bolometric fit compares the 32-phase pulse profile from the best-fitting two-spot model to the corresponding synthetic bolometric NICER waveform. This yields $\chi^2 = 29.85$. Following Section 4.3 of \citet{2024ApJ...974..295D}, we adopt an effective number of approximately primary parameters influencing the bolometric waveform. Including the overall rotational phase and a single unmodulated background term gives approximately $32 - 5 - 2 = 25$ dof, resulting in $\chi^2/dof = 29.85/25$. The corresponding probability of obtaining such a value or larger is $p = 0.23$, which is again acceptable.

As shown in Figure \ref{fig:main1}, the inferred mass and radius credible regions encompass the input values used to generate the synthetic data, with the radius posteriors (Table \ref{table:totresultsR}) showing no apparent bias. The full posterior distribution for this case is displayed in Figure \ref{fig:fullpt1} in the \hyperref[sec:App1]{Appendix}.

\subsection{The synthetic data include an additional background but the fitted model does not}\label{subsec:2}

In Case 2, the synthetic XMM-Newton data contained additional background beyond the blank-sky background, but the fitted model assumed only the blank-sky background. The meanings of the line and symbol types are the same as in Figure~\ref{fig:main1}. As shown in Figure~\ref{fig:main2}, the resulting radius posterior extends to smaller radii than the true value, producing a bias of roughly $1\sigma$ relative to the radius assumed in generating the synthetic data (see Table~\ref{table:totresultsR}). This is measurable, but still within the $90\%$ credible region.

This bias arises because neglecting the additional background in the model causes the inferred spot contribution to be overestimated, which reduces the inferred modulation amplitude. In turn, this yields an overestimate of the stellar compactness and, given the precise external mass constraint, an underestimate of the radius. The full posterior distribution is shown in Figure \ref{fig:fullpt2} in the \hyperref[sec:App1]{Appendix}.

The fits remain statistically acceptable. The $\chi^2/dof$ values for the phase–energy and bolometric fits are $2980.24/2901$ and $29.01/25$, with corresponding probabilities of 0.149 and 0.263, respectively. However, the Bayesian evidence for this model is lower than in Case 1 by $\Delta \log Z = -8.59 \pm 0.11$ (see Table~\ref{table:results}), indicating that the model without an additional background is less favored by a factor of $e^{8.59} \sim 5,400$. This comparison supports the conclusion that a waveform model including extra background provides a more accurate description of the data.

\subsection{The Synthetic Data Do Not Include an Additional Background but the Fitted Model Allows an Additional Background}\label{subsec:3}

In Case 3, we analyzed synthetic data that included only the XMM-Newton blank-sky background, but fit it with a model that allowed an additional background. When the model allows for additional background, the fit infers a larger unmodulated XMM background than was used to generate the synthetic data. Given the fixed synthetic data, the best-fit spot waveform will therefore underestimate the total number of counts that come from the spots. In general, some of the counts from the spots will be modulated, and some can be unmodulated. The NICER data unambiguously tell us the number of modulated counts, which must come from the hot spots, so the ratio of modulated counts to the total apparent spot counts becomes artificially high. Thus, the modulation fraction (i.e., the number of modulated counts divided by the inferred total counts coming from the spots) will appear to be larger than it should be. A fractional modulation that is too high leads to a larger inferred value of the inverse compactness (R/M). Given the mass constraint, the radius posterior extends to values larger than the radius that was assumed when the synthetic data were generated, producing a bias of about $1\sigma$ (see Figure~\ref{fig:main3}; the meanings of the line and symbol types are the same as in Figure~\ref{fig:main1}). The full posterior distribution is shown in Figure~\ref{fig:fullpt3} in the \hyperref[sec:App1]{Appendix}.

The fit quality for this model is statistically acceptable: the phase-channel $\chi^2/{\rm dof}$ is 2920.28/2901 ($p = 0.397$), and the bolometric $\chi^2/{\rm dof}$ is 39.24/25 ($p = 0.035$) (see Table \ref{table:results}). The log evidence for this fit is $\Delta \rm log\ Z = -5.76 \pm 0.10$ relative to the Case 4 model (which assumes only the blank-sky background), indicating that the model with an additional background is disfavored by a factor of $e^{5.76}\sim300$. This suggests that including an additional background component when the data do not contain one is disfavored, which may reflect both the absence of a corresponding improvement in fit quality and the penalty associated with the additional parameter volume.

\subsection{The Synthetic Data Does Not Include and the Fitted Model Does Not Assume an Additional Background}\label{subsec:4}

In Case~4 we fit a model that assumes the XMM-Newton blank-sky background is the only non-spot background to a synthetic pulse waveform that was generated using a model that made this same assumption. In this case, the radius estimate agrees with the radius assumed in the model (see Figure~\ref{fig:main4} and Table~\ref{table:totresultsR}). The phase-channel $\chi^2/{\rm dof}$ is 2916.59/2901, which has a probability of 0.416, and the bolometric $\chi^2/{\rm dof}$ is 37.41/25, which has a probability of 0.053 (see Table~\ref{table:results}). Both probabilities are acceptable. The full posterior distribution is shown in Figure~\ref{fig:fullpt4} in the \hyperref[sec:App1]{Appendix}. 

\hspace{0.01cm}

\section{Summary and Conclusions}\label{sec:conc.}

In this work, we have investigated how the assumption made about the X-ray background of nonaccreting X-ray pulsars when fitting pulse waveform models jointly to NICER and XMM-Newton can affect estimates of the radii of these neutron stars, focusing on the potential for systematic error. We considered the four combinations that occur if one assumes that the background in the data is or is not solely the XMM-Newton blank-sky background and that the background assumed in the model is or is not solely the XMM-Newton background.

If the data contain a background (produced, e.g., by cosmic rays, AGN, or unresolved X-ray sources) that is missing from the model, the inferred stellar radius decreases. This is because a more compact star produces greater spacetime curvature and thus weaker modulation, mimicking the effect of a larger background. Conversely, if the model includes a background that is not present in the data, the inferred radius increases to preserve the fractional modulation. The radius estimate is unbiased compared to the actual stellar radius when the background in the model matches the background in the data.

We considered a large additional background in one of our synthetic data sets, allowing a background equal to the full smoothed synthetic count rate given by the pointed XMM-Newton observation. Even with this very large unmodeled background, the bias in the estimated stellar radius was small, corresponding to only about a $\sim1\sigma$ offset from the true value.  Also, in every case we considered, the Bayesian evidence favored the correct model, which suggests its usefulness in choosing the best model for analyzing this kind of data. This was so even though our investigation considered only synthetic pulse waveform data similar to the NICER data from PSR~J0740$+$6620, which has a relatively low signal-to-noise ratio.

These results add to the evidence that NICER-like analyses provide accurate measurements of neutron star radii as long as the statistical sampling is thorough and the data are fit well by the model.

\section*{Acknowledgments} \label{sec:Ack}

This work was supported in part by NASA ADAP grant 80NSSC21K0649, and was supported by NASA through the NICER mission and the Astrophysics Explorers Program. The authors acknowledge the University of Maryland supercomputing resources (http://hpcc.umd.edu) that were made available for conducting the research reported in this paper. This work also utilized computational resources provided by the NASA Center for Climate Simulation (NCCS) Discover supercomputing facility. We are grateful to the NICER team for their ongoing efforts in instrument operation, data calibration, and analysis support. This work was performed in part at the Aspen Center for Physics, which is supported by National Science Foundation grant PHY-2210452. 
A.J.D was supported by NASA through the NASA Hubble Fellowship grant No. HST-HF2-51553.001, awarded by the Space Telescope Science Institute, which is operated by the Association of Universities for Research in Astronomy, Inc., for NASA, under contract NAS5-26555. We also thank Gokul Srinivasragavan for valuable feedback and assistance in reviewing an early draft of this manuscript.

\software{\texttt{emcee} \citep{2013PASP..125..306F},
          \texttt{Python} and \texttt{NumPy} \citep{2007CSE.....9c..10O},
          \texttt{Matplotlib} \citep{2007CSE.....9...90H},
          \texttt{Cython} \citep{2011CSE....13b..31B},
          \texttt{ptemcee} \citep{2021ascl.soft01006V},
          \texttt{MultiNest} \citep{2009MNRAS.398.1601F},
          \texttt{schwimmbad} \citep{2017JOSS....2..357P},
          \texttt{HEASoft} \cite{2014ascl.soft08004N},
          \texttt{pocoMC} \citep{2022JOSS....7.4634K}}

\bibliography{sample631}{}
\bibliographystyle{aasjournal}

\section{Appendix: Posterior Distributions from Analyses of Synthetic J0740-like NICER and XMM-Newton Data Using \texttt{MultiNest} and \texttt{ptemcee}} \label{sec:App1}

\begin{figure}
      \includegraphics[width=\linewidth]{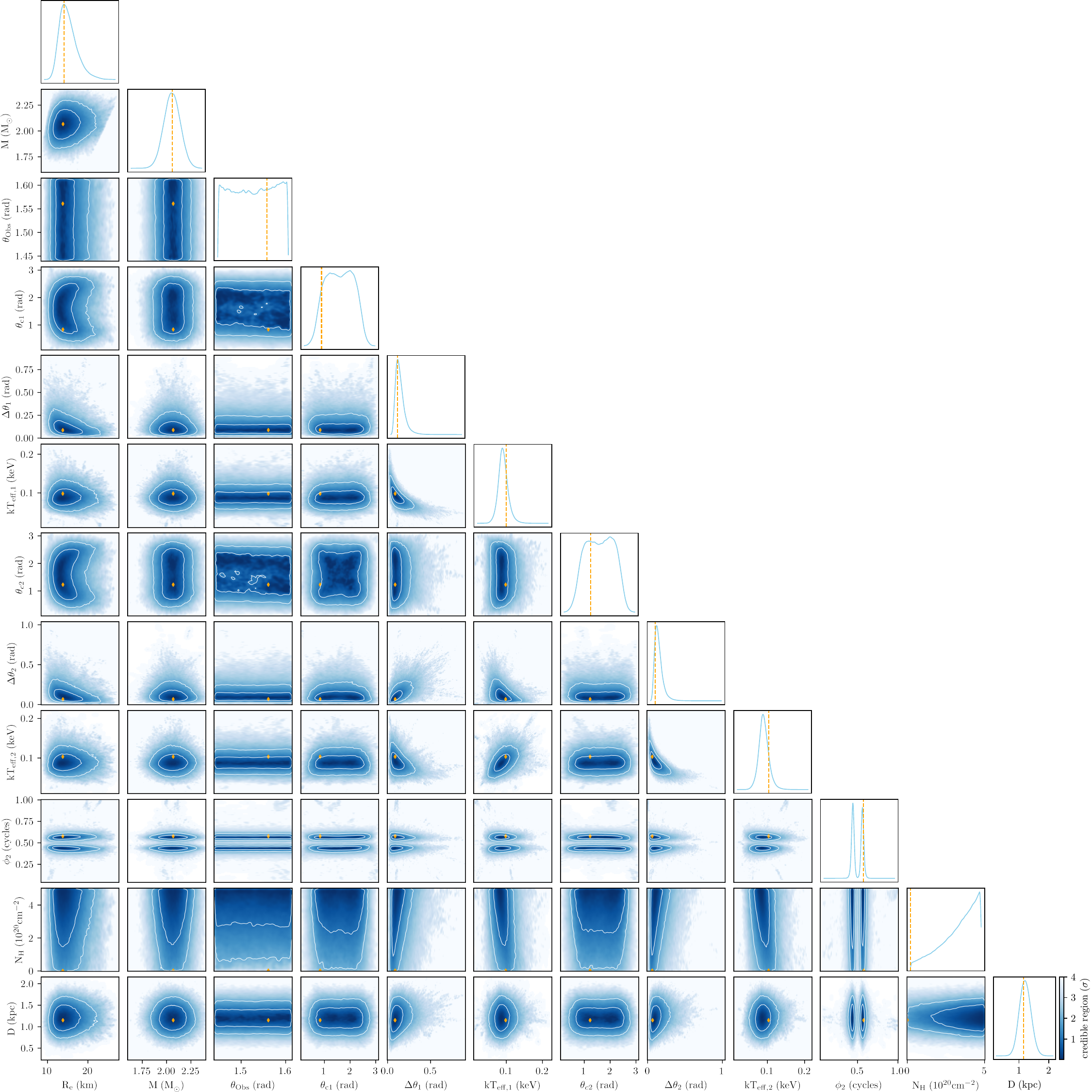}
      \caption{The full set of 1D and 2D posterior distributions from our \texttt{MultiNest} and \texttt{ptemcee} analysis of Case~1. The synthetic data were constructed assuming that the background is the XMM-Newton blank-sky background plus an additional background (see text). The analysis model also assumes that the background is the XMM-Newton blank-sky background plus an additional background.}
      \label{fig:fullpt1}
\end{figure}

\begin{figure}
      \includegraphics[width=\linewidth]{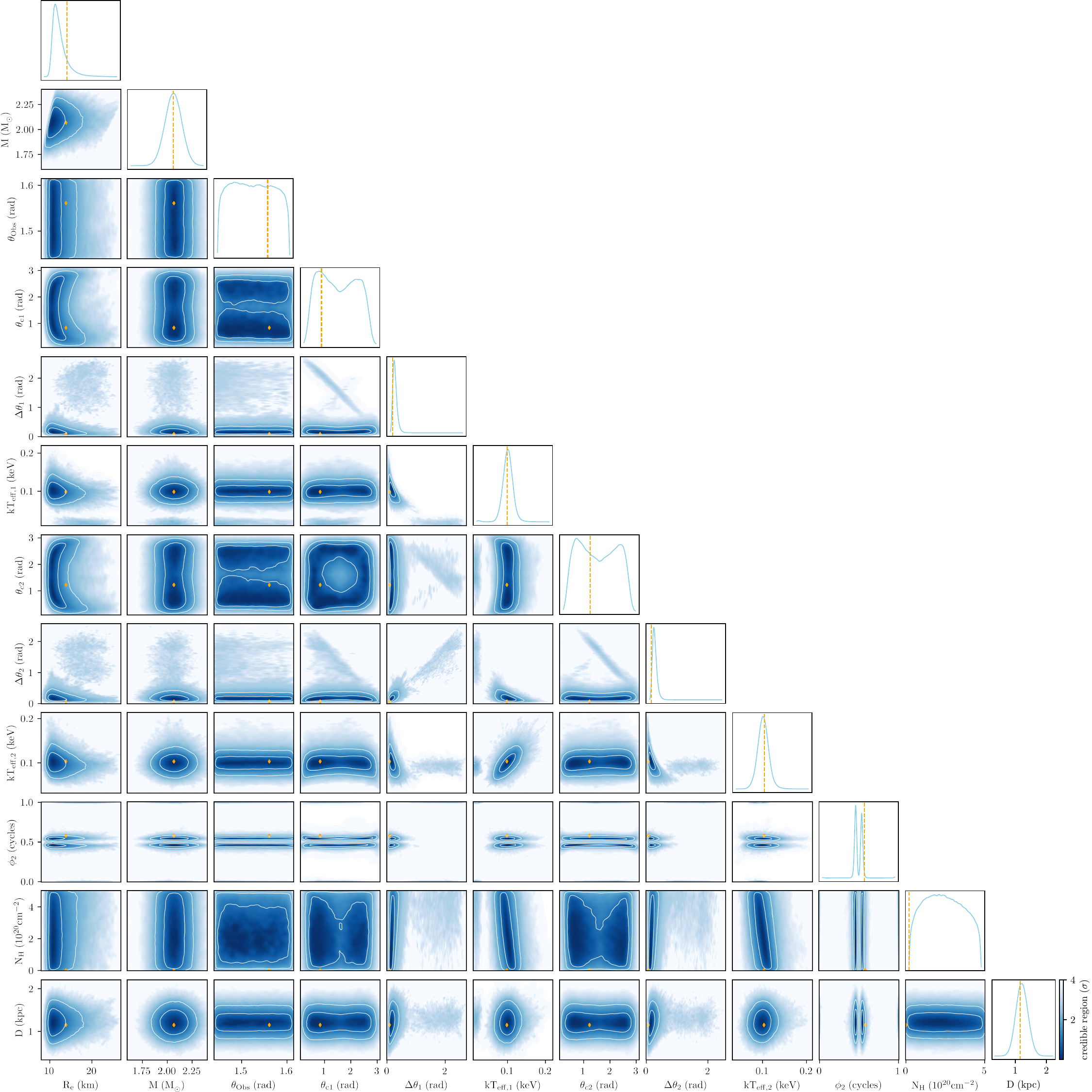}
      \caption{The full set of 1D and 2D posterior distributions from our \texttt{MultiNest} and \texttt{ptemcee} analysis of Case~2. The synthetic data were constructed assuming that the background is the XMM-Newton blank-sky background plus an additional background (see text). The analysis model assumes that the XMM-Newton blank-sky background is the only background.}
      \label{fig:fullpt2}
\end{figure}

\begin{figure}
      \includegraphics[width=\linewidth]{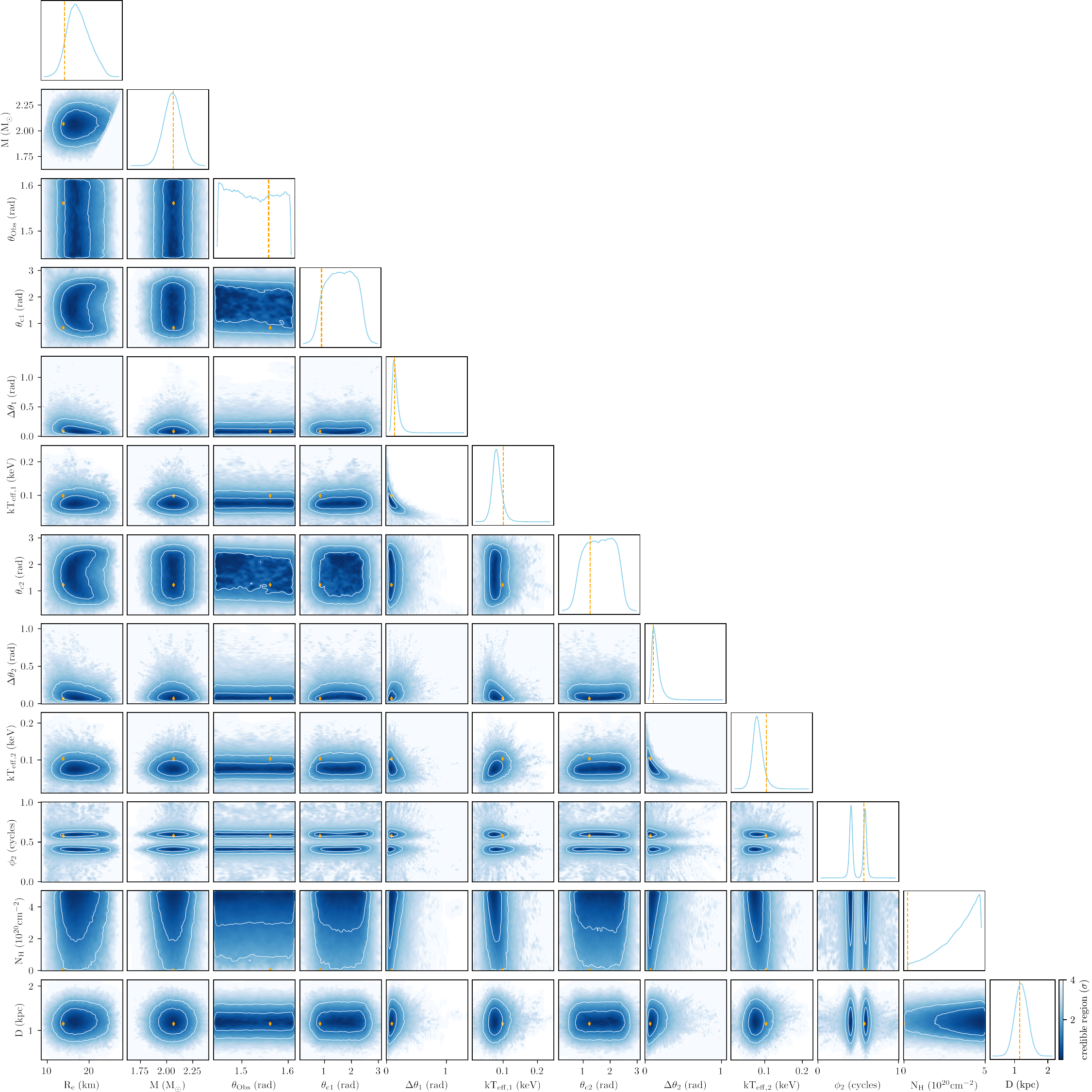}
      \caption{The full set of 1D and 2D posterior distributions from our \texttt{MultiNest} and \texttt{ptemcee} analysis of Case~3. The synthetic data were constructed assuming that the background is the XMM-Newton blank-sky background. The analysis model assumes that the background is the XMM-Newton blank-sky background plus an additional background.}
      \label{fig:fullpt3}
\end{figure}

\begin{figure}
      \includegraphics[width=\linewidth]{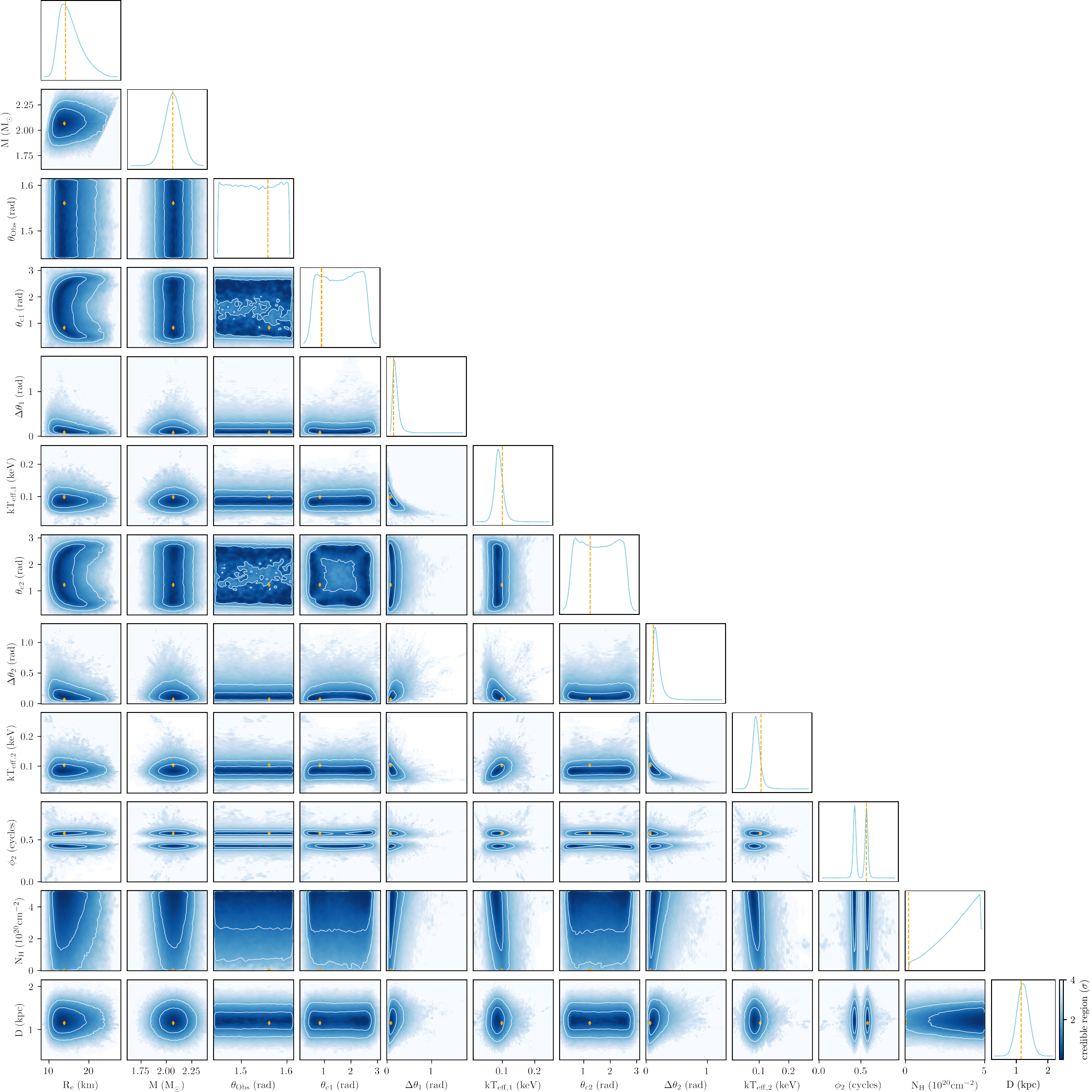}
      \caption{The full set of 1D and 2D posterior distributions from our \texttt{MultiNest} and \texttt{ptemcee} analysis of Case~4. The synthetic data and the model both assume that the XMM-Newton blank-sky background is the only background.}
      \label{fig:fullpt4}
\end{figure}

Figures \ref{fig:fullpt1}–\ref{fig:fullpt4} show the full set of posterior distributions for the neutron star radii together with other model parameters that are not included in Figures \ref{fig:main1}–\ref{fig:main4} of the main text.

\end{document}